\documentclass[%
 reprint,
 amsmath,amssymb,
 aps,
]{revtex4-1}

\usepackage{graphicx}
\usepackage{dcolumn}
\usepackage{bm}

\begin{document}


\title{Lorentz TEM investigation of chiral spin textures and N\'eel Skyrmions in asymmetric [Pt/(Co/Ni)$_M$/Ir]$_N$ multi-layer thin films}

\author{Maxwell Li}
\email{mpli@andrew.cmu.edu}
\author{Derek Lau}
\author{Marc De Graef}
\author{Vincent Sokalski}

\affiliation{%
 Department of Materials Science \& Engineering, Carnegie Mellon University, Pittsburgh, Pennsylvania 15213, USA
}%

\date{\today}

\begin{abstract}
We examine magnetic domain patterns in symmetric [Co/Ni]$_M$ and asymmetric [Pt/(Co/Ni)$_M$/Ir]$_N$ multi-layers using Fresnel mode Lorentz transmission electron microscopy (LTEM). In the symmetric multi-layer, where the Dzyaloshinskii-Moriya Interaction is expected to be zero, we observe purely Bloch type domain walls with no preferred chirality. In the asymmetric multi-layers, where significant interfacial DMI is present, we observe domain patterns with chiral N\'eel domain walls, which evolve into sub-100nm isolated N\'eel Skyrmions with the application of a perpendicular field.  The impact of layer thickness and film stack on interfacial magnetic properties is discussed in the context of developing a tunable multi-layer system for future spintronic applications.
 
\begin{description}

\item[DOI]
Secondary publications and information retrieval purposes.
\end{description}
\end{abstract}

\pacs{Valid PACS appear here}
\maketitle

\section{Introduction}
Magnetic objects with added topological stability, such as Skyrmions and chiral domain walls (DWs), have garnered a great deal of attention in recent years due the unprecedented efficiency by which they can be manipulated with electric current for use in future spintronic devices \cite{Bromberg2014,Emori2013,Ryu2014}. The topology of such objects is described by the topological charge, $C$, as determined from $4\pi\,C=\int \mathbf{m} \cdot \left( \partial_x \mathbf{m} \times \partial_y \mathbf{m} \right) \mathrm{d}x\mathrm{d}y$. Such objects are stabilized by the Dzyaloshinskii-Moriya interaction (DMI) which is found in magnetic materials where inversion symmetry is broken \cite{Dzyaloshinsky1958,Moriya1960}. This interaction has been observed in bulk magnetic materials lacking inversion symmetry (such as FeGe \cite{Yu2011} and MnSi \cite{Yu2015}) and more recently in magnetic multi-layers at the interface of a ferromagnet (FM) and a heavy metal (HM) with large spin-orbit coupling \cite{Thiaville2012,Hrabec2014,Pellegren2017}. In the latter case, interfacial DMI is well-established to stabilize chiral N\'eel DWs over the magnetostatically favorable Bloch wall \cite{Thiaville1995}.

The discovery of DMI has opened the door to a range of new magnetic materials and multi-layer designs to support the formation of such chiral configurations. Parameters such as FM/HM interfaces \cite{Chen2013,Yang2018,Khadka2018}, FM layer composition \cite{Soumy2017}, and asymmetric stacking sequences  \cite{Franken2014} have been explored as means of strengthening DMI in multi-layer systems. The ideal system will offer tunability of emergent magnetic properties, including DMI, while preserving other critical properties such as magnetic anisotropy and saturation magnetization.

In this work we examine the magnetic domain structures of asymmetric multi-layers based on [Pt/(Co/Ni)$_M$/Ir]$_N$ using Lorentz transmission electron microscopy (LTEM). In these films, interfacial DMI is induced at the interface of Pt/Co and Ni/Ir. It has been reported that Pt and Ir induce DMI of opposite sign leading to an additive effect when placed on opposite surfaces of a magnetic heterostructure \cite{Chen2013,MoreauLuchaire2016}. Although some reports have found both the Pt/Co and Ir/Co interfaces to produce DMI of the same sign\cite{Ryu2014,Lau2018}, it is widely agreed that the magnitude of DMI at the Pt/Co interface is much larger.  To estimate the strength of DMI in films examined here, we have leveraged the asymmetric bubble expansion technique using Kerr microscopy \cite{je2013,Hrabec2014,lavrijsen2015} for $M = 2$ and $N = 1$, as shown in the supplemental information.\cite{SupMat}  In films used for the Lorentz TEM investigation presented here, the number of Co/Ni layers, $M$, offers tunability of interfacial magnetic properties, such as DMI, while the number of total repeats, $N$, contributes to the stabilization of stripe domain patterns \cite{Kooy1960,Cape1971}. These characteristics are reflected in Fresnel-mode LTEM images where N\'eel walls are observed confirming the presence of interfacial DMI in our materials system.  We note that the inclusion of Ni in the multi-layer allows us to increase the magnetic layer thickness (via $M$), which allows variation in DMI, but preserves perpendicular magnetic anisotropy due to the Co/Ni interface. This would not be possible in a multi-layer based only on Pt/Co/Ir.  Symmetric [Co/Ni]$_M$ based multi-layers were also examined with LTEM for comparison; they are expected to have zero DMI and display only Bloch domain walls.

\section{Experimental}

Multi-layers were deposited onto $5$ nm thick SiN membranes via magnetron sputtering in an Ar environment with working pressure fixed at $2.5$ mTorr and base pressure at $<3.0\times 10^{-7}$ Torr. All film stacks were deposited onto seedlayers of Ta(3)/Pt(3) with units in nm. Asymmetric and symmetric multi-layers had the following subsequent layers: [Pt(0.5)/(Co(0.2)/Ni(0.6))$_M$/Ir(0.5)]$_N$ and [Co(0.2)/Ni(0.6)]$_M$/Co(0.2), respectively.  Magnetic hysteresis loops of these multi-layers were measured using alternating gradient field magnetometry (AGFM). 

Fresnel-mode LTEM imaging was performed on an aberration-corrected FEI Titan G2 80-300 operated in Lorentz mode. Magnetic induction maps were produced by solving the Transport of Intensity Equation (TIE) using over- and under-focused Fresnel images as in \cite{phatak2016}. In some cases, a perpendicular magnetic field was applied in-situ by exciting the objective lens of the microscope. Interfacial DMI affects domain wall characteristics most notably manifest in the formation of N\'eel walls over the magnetostatically favorable Bloch wall \cite{Thiaville1995}. In the absence of specimen tilt, N\'eel walls do not display magnetic contrast as the deflection of electrons through the Lorentz force lies parallel to the DW. When a sample tilt is applied, however, the perpendicular magnetic induction of surrounding domains gains an effective in-plane induction which deflects electrons towards or away from the DW, leading to the appearance of magnetic contrast \cite{Benitez2015,McVitie2018}. This is not the case for Bloch walls which form magnetic contrast in the absence of sample tilt as electrons are deflected perpendicular to the DW.

\begin{figure}[b]
    \includegraphics{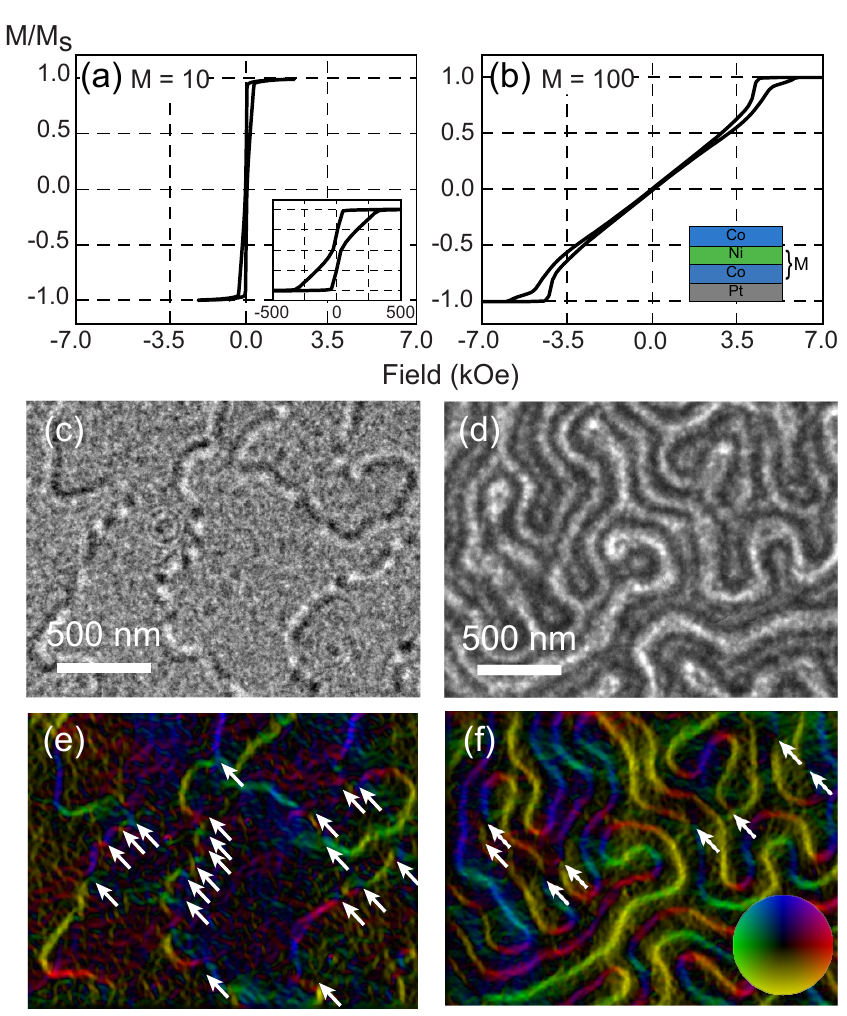}
    \caption{\small Perpendicular M-H loops, Fresnel mode image, and calculated in-plane induction maps of symmetric [Co/Ni]$_M$ multi-layers where $M =10 $ (a,c,e) and $100$ (b,d,e). Arrows designate locations of vertical Bloch lines in in-plane induction maps. Inset depicts $M = 10$ loop with a smaller field of view and schematic of symmetric film stack.}.
    \label{fig:symmetricMH}
\end{figure}

\section{Symmetric Co/Ni Multi-Layers}

 Symmetric [Co/Ni]$_M$ based multi-layers were first examined to serve as a limiting case where there is no impact from Ir and Pt. Perpendicular M-H loops from these films indicate perpendicular magnetic anisotropy is present.  The shearing and pinched shape of the loop, which is characteristic of bubble materials, suggests the formation of a multi-domain state at zero applied field.\cite{Kooy1960,Cape1971}  The increased loop shearing at $M = 100$, compared to $M = 10$, is due to the increased role of dipole-dipole interactions in the formation of small magnetic domains when films are thicker.  We note a negative nucleation field of 4 kOe for $M = 100$ as compared to 50 Oe for $M = 10$.  

Fresnel mode LTEM images of both symmetric multi-layers (M = 10, 100) display magnetic contrast in the absence of tilt (FIG. \ref{fig:symmetricMH}c \& d); the presence of these Bloch walls indicates no DMI is present in these multi-layers as expected. The domain structure of both multi-layers depict a demagnitized labyrinth-configuration, as reflected in M-H loops, with $M = 100$ displaying uniform domain widths throughout the field of view whereas those in $M = 10$ are less periodic.  

\begin{figure}[b]
    \includegraphics{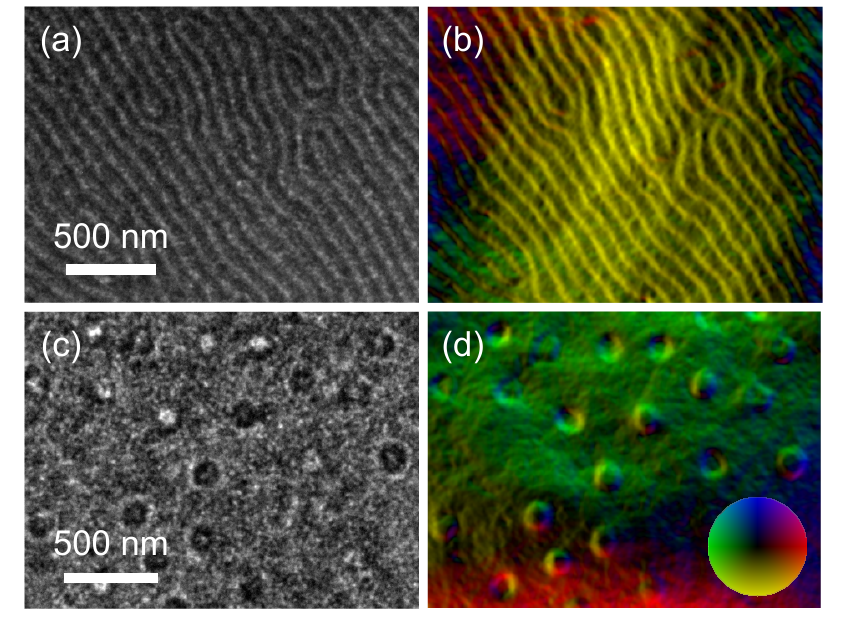}
    \caption{\small Fresnel mode LTEM images (a,c) and respective in-plane induction maps (b,d) of symmetric [Co/Ni]$_{100}$ multi-layers a,b) after ex-situ in-plane saturation and c,d) with static in-situ perpendicular field, $H_z = 4800 Oe$. }.
    \label{fig:symmetric}
\end{figure}

Vertical Bloch lines (VBLs) are also observed in these multi-layers which are decribed by 180$^{\circ}$ rotations in magnetic induction along a domain wall.\cite{malozemoff1972} This appears as a discontinuity in contrast along a DW in Fresnel mode images whereby the contrast inverts about the discontinuity. Such VBLs are observed to occur at a high frequency in the $M = 10$ multi-layers forming clusters whereby several VBLs exist at close proximity with one another along a DW.\cite{zvezdin1986} VBLs are also observed in $M = 100$ multi-layers but are harder to discern as the DWs are spaced closer together than those in $M = 10$. Application of an ex-situ in-plane magnetic field produces a stripe domain pattern with domains aligned parallel to the field direction as shown in FIG. \ref{fig:symmetric}a,b.  An in-situ perpendicular magnetic field leads to the formation of magnetic bubbles, which are found to be Bloch type (C = 1) having no preferred chirality or topologically trivial (C = 0) where two VBLs are present along the circumference (see example in supplementary information).\cite{SupMat}

\section{Asymmetric Pt/Co/Ni/Ir Multi-Layers}

\begin{figure}[b]
    \includegraphics{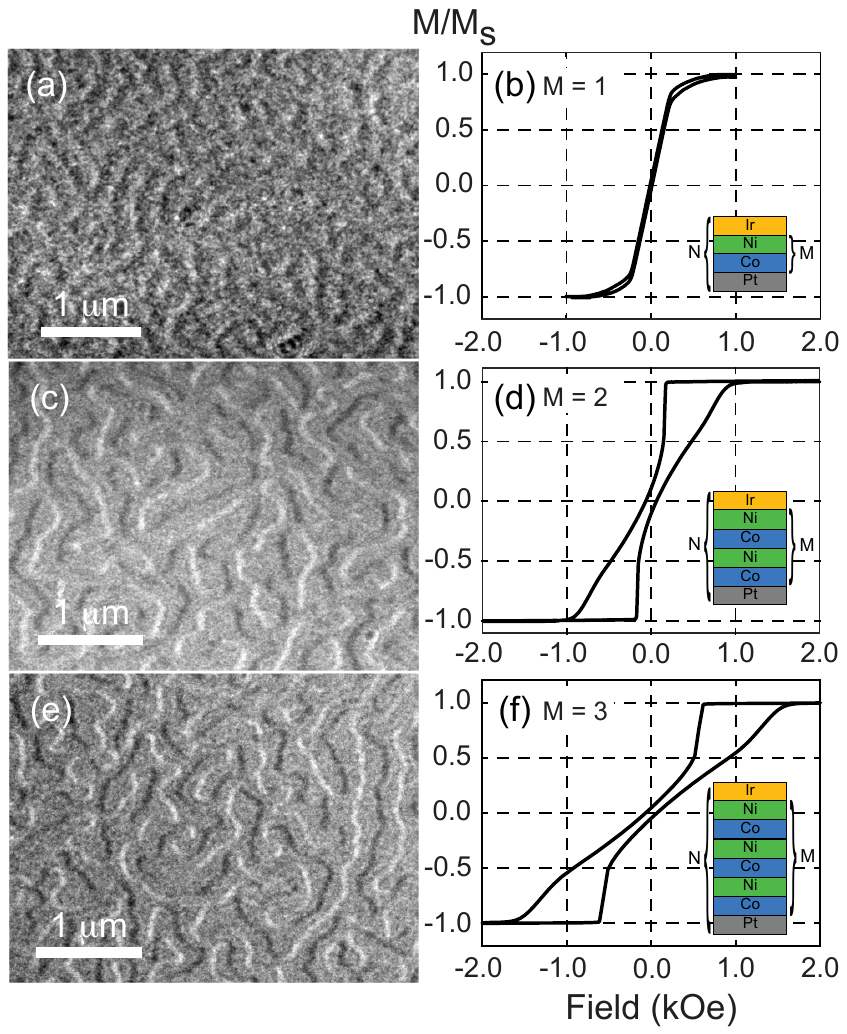}
    \caption{\small Fresnel mode Lorentz TEM images and perpendicular M-H loops of asymmetric asymmetric [Pt/(Co/Ni)$_M$/Ir]$_{10}$ multi-layers where M = (a,b) 1, (c,d) 2, or (e,f) 3. Sample tilt of 20$^\circ$ is present in each image. Insets depict schematic of asymmetric film stack.}.
    \label{fig:AsymmCompare}
\end{figure}

The tunability of asymmetric [Pt/(Co/Ni)$_M$/Ir]$_N$ based multi-layers was examined by varying the number of Co/Ni layers in a repeat unit, $M$. The total number of repeat units in a multi-layer stack, $N$, was used to stabilize labyrinth domain patterns. By increasing $M$, the effective DMI is expected to decrease while the areal magnetization will increase.  We note that the interface anisotropy associated with Co/Ni is comparable to that of Co/(Pt, Ir) so we do not expect significant change in perpendicular magnetic anisotropy (PMA).  We note that increasing thickness of either Co or Ni would lead to a decrease in PMA, which could be useful for stabilization of Skyrmions in subsequent studies. Broadly, this tunability of DMI without compromising perpendicular magnetic anisotropy would allow for a more comprehensive examination of chiral magnetic textures in thin film systems.

M-H loops of asymmetric multi-layers (FIG. \ref{fig:AsymmCompare}) indicate perpendicular magnetic anisotropy is preserved with the addition of heavy metal layers in the repeat structure. Like the symmetric case, the loop becomes more sheared with increasing $M$ and has a similar pinched hysteresis.  

\begin{figure}[t]
    \includegraphics{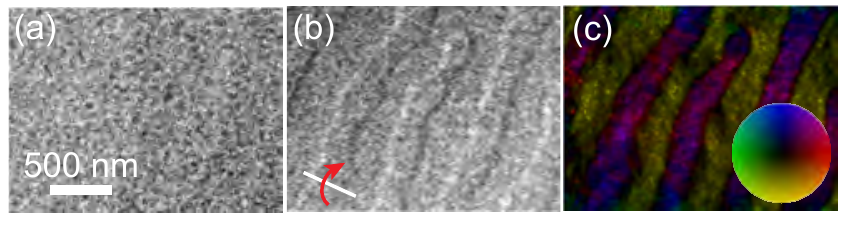}
    \caption{\small Fresnel mode Lorentz TEM images of [Pt/(Co/Ni)$_2$/Ir]$_{5}$ (a) for no tilt and (b) 20$^\circ$ tilt. (c) In-plane 
    induction map calculated from tilted through-focus images.}.
    \label{fig:asymminduction}
\end{figure}

Fresnel mode LTEM images of these asymmetric multi-layers displayed no magnetic contrast in the non-tilted state (FIG. \ref{fig:asymminduction}). Upon application of a 20$^\circ$ tilt, however, magnetic contrast becomes apparent indicating the presence of N\'eel walls. The contrast seen in the accompanied in-plane induction map calculated from the tilted Fresnel images is due to the magnetization of the domains, which now have a component perpendicular to the electron beam.  This process reveals alternating perpendicular domains, but does not directly provide any details on the internal structure of the domain walls.  As such we proceed on characterizing the domain patterns in asymmetric films by direct examination of the Fresnel images. Magnetic contrast characteristic of N\'eel walls is observed even when $M$ is increased from 1 to 3 repeats despite an effective reduction in DMI. In first approximation, due to the changing magnetic layer thickness, we expect DMI of -0.772, -0.386, and -0.257 mJ/m$^2$ for $M$ = 1, 2, and 3, respectively, based on the experimental measurements of asymmetric bubble expansion shown in the supplemental information for $M$ = 2.\cite{SupMat} These values are large enough to overcome the DW anisotropy and produce pure N\'eel DWs.  A labyrinth domain structure is observed in each of these multi-layers with DW contrast becoming more apparent as $M$ increases due to greater magnetic induction originating from a larger ferromagnetic thickness. Despite changes to DMI and perpendicular magnetic anisotropy, the domain widths are not observed to change greatly between $M$ = 2 and 3. When compared with symmetric films, the domain widths observed in these asymmetric films are much smaller which stems from a reduction in the DW energy due to DMI.

\begin{figure}[t]
    \includegraphics{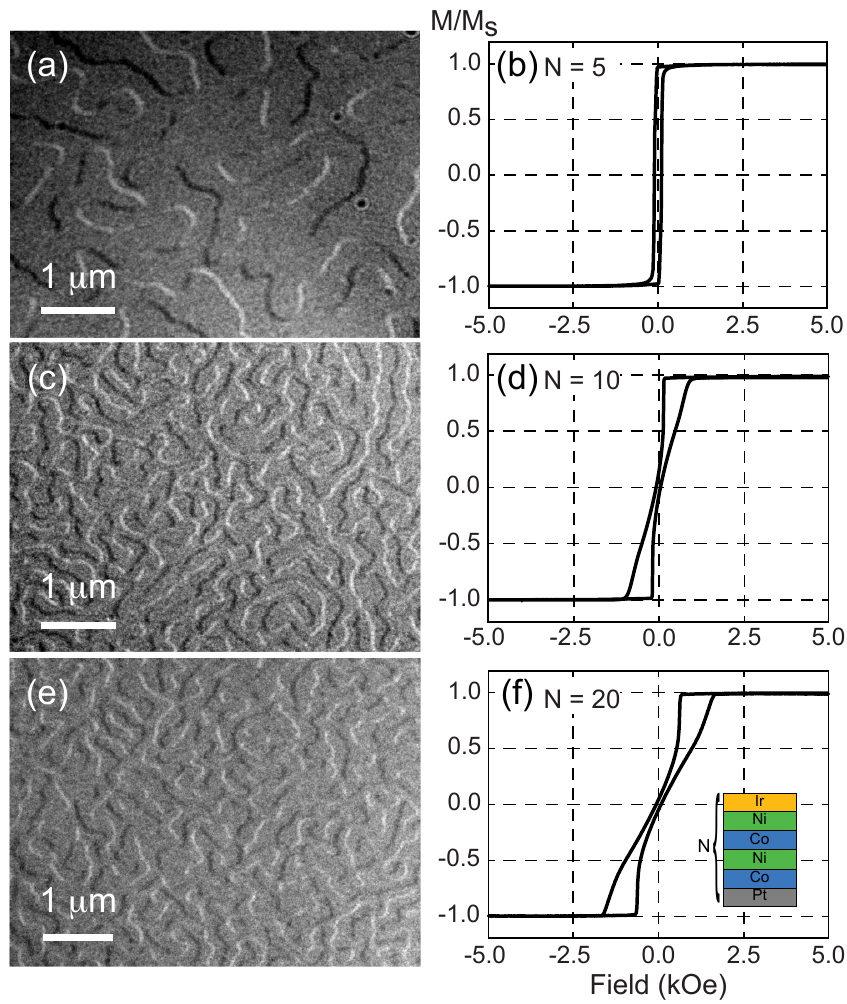}
    \caption{\small Fresnel mode Lorentz TEM images and perpendicular M-H loops of asymmetric [Pt/(Co/Ni)$_2$/Ir]$_{N}$ where $N =$ (a,b) 5, (c,d) 10, and (e,f) 20. Samples were tilted by 20$^\circ$ for each image. Inset depicts schematic of asymmetric film stack.}.
    \label{fig:CoNi2Comapre}
\end{figure}

\begin{figure}[b]
    \includegraphics{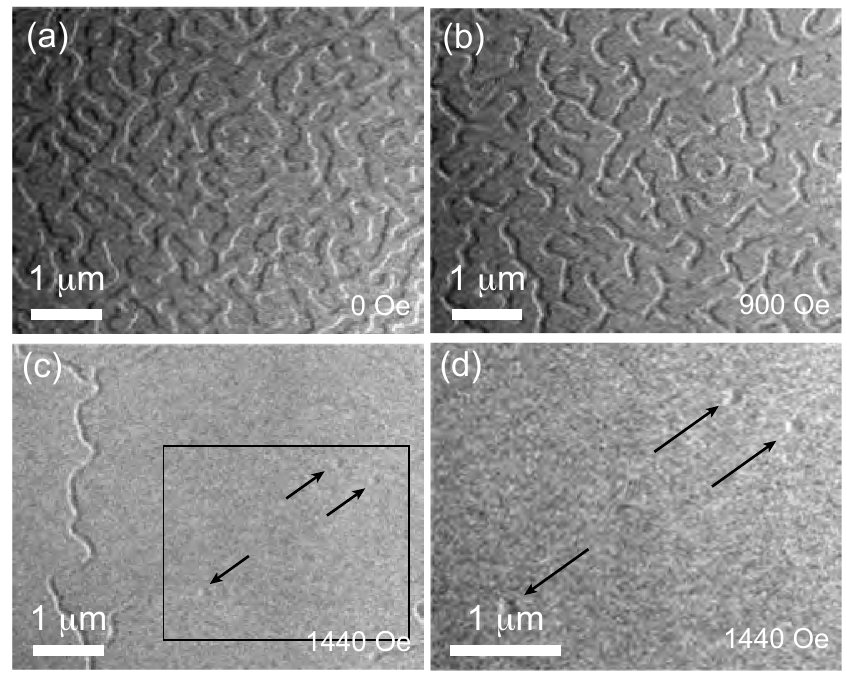}
    \caption{\small Fresnel mode Lorentz TEM images of asymmetric [Pt/(Co/Ni)$_2$/Ir]$_{20}$ with increasing perpendicular magnetic field applied in-situ. Arrows indicate positions of N\'eel Skyrmions in c) and d). Samples were tilted by 20$^\circ$ for each image.}.
    \label{fig:HZ}
\end{figure}

Next we examined the effects the number of film stack repeats, $N$, has on magnetic domain characteristics in [Pt/(Co/Ni)$_2$/Ir]$_N$ where $N =$ 5, 10 or 20 (FIG \ref{fig:CoNi2Comapre}). Magnetic domains when $N$ = 5 were noticeably wider than those for $N =$ 10 and 20 which is also observed with symmetric films. A perpendicular magnetic field was applied in-situ on [Pt/(Co/Ni)$_2$/Ir]$_{20}$ multi-layers by exciting the objective lens of the TEM (FIG. \ref{fig:HZ}). With increasing field, domains with magnetization anti-parallel to the direction of the field shrink and visa versa. Near the saturation field, these domains form isolated N\'eel Skyrmions with diameters of $\sim$80 nm (FIG \ref{fig:HZ}c) before annihilating.  The inversion of contrast upon reversal of focus confirms the magnetic origin of these features (see supplementary info). Skyrmions were not observed to form in [Pt/(Co/Ni)$_2$/Ir]$_{N}$ multi-layers where $N =$ 5 or 10 and instead formed long, worm-like domains before annihilating at the saturation field (see supplementary info). 

\section{Summary}
In summary, we have examined asymmetric Pt/Co/Ni/Ir based multi-layers using Lorentz transmission electron microscopy. The properties of these multi-layers were tuned through variations in ferromagnetic layer thickness and overall thickness, which were reflected in the magnetic domain structure. Symmetric Co/Ni multi-layers displayed only Bloch walls indicating no DMI was present; with the addition of a Pt and Ir layer sandwiching Co and Ni, DMI is induced which is reflected in the presence of exclusively N\'eel walls in Fresnel mode images. Although the effective DMI was diminished with greater ferromagnetic content in film stack repeats, N\'eel walls were still observed. Additionally, asymmetric Pt/Co/Ni/Ir multi-layers with greater number of total film stack repeats were observed to support the formation of sub-100 nm Skyrmions at room temperature in the presence of a perpendicular magnetic field. Overall, this materials system provides a tunable platform for further exploration of chiral spin textures and the development of spintronic devices.       

\section*{Acknowledgements}
This work is financially  supported by the Defense Advanced Research Project Agency (DARPA) program on Topological Excitations in Electronics (TEE) under grant number D18AP00011. The authors also acknowledge use of the Materials Characterization Facility at Carnegie Mellon University supported by grant MCF-677785.

\bibliography{citations}

\begin{thebibliography}{29}%
\makeatletter
\providecommand \@ifxundefined [1]{%
 \@ifx{#1\undefined}
}%
\providecommand \@ifnum [1]{%
 \ifnum #1\expandafter \@firstoftwo
 \else \expandafter \@secondoftwo
 \fi
}%
\providecommand \@ifx [1]{%
 \ifx #1\expandafter \@firstoftwo
 \else \expandafter \@secondoftwo
 \fi
}%
\providecommand \natexlab [1]{#1}%
\providecommand \enquote  [1]{``#1''}%
\providecommand \bibnamefont  [1]{#1}%
\providecommand \bibfnamefont [1]{#1}%
\providecommand \citenamefont [1]{#1}%
\providecommand \href@noop [0]{\@secondoftwo}%
\providecommand \href [0]{\begingroup \@sanitize@url \@href}%
\providecommand \@href[1]{\@@startlink{#1}\@@href}%
\providecommand \@@href[1]{\endgroup#1\@@endlink}%
\providecommand \@sanitize@url [0]{\catcode `\\12\catcode `\$12\catcode
  `\&12\catcode `\#12\catcode `\^12\catcode `\_12\catcode `\%12\relax}%
\providecommand \@@startlink[1]{}%
\providecommand \@@endlink[0]{}%
\providecommand \url  [0]{\begingroup\@sanitize@url \@url }%
\providecommand \@url [1]{\endgroup\@href {#1}{\urlprefix }}%
\providecommand \urlprefix  [0]{URL }%
\providecommand \Eprint [0]{\href }%
\providecommand \doibase [0]{http://dx.doi.org/}%
\providecommand \selectlanguage [0]{\@gobble}%
\providecommand \bibinfo  [0]{\@secondoftwo}%
\providecommand \bibfield  [0]{\@secondoftwo}%
\providecommand \translation [1]{[#1]}%
\providecommand \BibitemOpen [0]{}%
\providecommand \bibitemStop [0]{}%
\providecommand \bibitemNoStop [0]{.\EOS\space}%
\providecommand \EOS [0]{\spacefactor3000\relax}%
\providecommand \BibitemShut  [1]{\csname bibitem#1\endcsname}%
\let\auto@bib@innerbib\@empty
\bibitem [{\citenamefont {Bromberg}\ \emph {et~al.}(2014)\citenamefont
  {Bromberg}, \citenamefont {Moneck}, \citenamefont {Sokalski}, \citenamefont
  {Zhu}, \citenamefont {Pileggi},\ and\ \citenamefont {Zhu}}]{Bromberg2014}%
  \BibitemOpen
  \bibfield  {author} {\bibinfo {author} {\bibfnamefont {D.~M.}\ \bibnamefont
  {Bromberg}}, \bibinfo {author} {\bibfnamefont {M.~T.}\ \bibnamefont
  {Moneck}}, \bibinfo {author} {\bibfnamefont {V.~M.}\ \bibnamefont
  {Sokalski}}, \bibinfo {author} {\bibfnamefont {J.}~\bibnamefont {Zhu}},
  \bibinfo {author} {\bibfnamefont {L.}~\bibnamefont {Pileggi}}, \ and\
  \bibinfo {author} {\bibfnamefont {J.-G.}\ \bibnamefont {Zhu}},\ }\href
  {\doibase 110.1109/IEDM.2014.7047159} {\bibfield  {journal} {\bibinfo
  {journal} {2014 IEEE International Electron Devices Meeting}\ ,\ \bibinfo
  {pages} {33.1.1}} (\bibinfo {year} {2014})}\BibitemShut {NoStop}%
\bibitem [{\citenamefont {Emori}\ \emph {et~al.}(2013)\citenamefont {Emori},
  \citenamefont {Bauer}, \citenamefont {Ahn}, \citenamefont {Martinez},\ and\
  \citenamefont {Beach}}]{Emori2013}%
  \BibitemOpen
  \bibfield  {author} {\bibinfo {author} {\bibfnamefont {S.}~\bibnamefont
  {Emori}}, \bibinfo {author} {\bibfnamefont {U.}~\bibnamefont {Bauer}},
  \bibinfo {author} {\bibfnamefont {S.-M.}\ \bibnamefont {Ahn}}, \bibinfo
  {author} {\bibfnamefont {E.}~\bibnamefont {Martinez}}, \ and\ \bibinfo
  {author} {\bibfnamefont {G.~S.~D.}\ \bibnamefont {Beach}},\ }\href {\doibase
  10.1038/nmat3675} {\bibfield  {journal} {\bibinfo  {journal} {Nature
  Materials}\ ,\ \bibinfo {pages} {pages 611}} (\bibinfo {year}
  {2013})}\BibitemShut {NoStop}%
\bibitem [{\citenamefont {Kwang-Su~Ryu}(2014)}]{Ryu2014}%
  \BibitemOpen
  \bibfield  {author} {\bibinfo {author} {\bibfnamefont {L.~T. . S. S. P.~P.}\
  \bibnamefont {Kwang-Su~Ryu}, \bibfnamefont {See-Hun~Yang}},\ }\href {\doibase
  10.1038/ncomms4910} {\bibfield  {journal} {\bibinfo  {journal} {Nature
  Communications}\ ,\ \bibinfo {pages} {3910}} (\bibinfo {year}
  {2014})}\BibitemShut {NoStop}%
\bibitem [{\citenamefont {Dzyaloshinsky}(1958)}]{Dzyaloshinsky1958}%
  \BibitemOpen
  \bibfield  {author} {\bibinfo {author} {\bibfnamefont {I.}~\bibnamefont
  {Dzyaloshinsky}},\ }\href@noop {} {\bibfield  {journal} {\bibinfo  {journal}
  {Journal of physics and chemistry of solids}\ }\textbf {\bibinfo {volume}
  {4}},\ \bibinfo {pages} {241} (\bibinfo {year} {1958})}\BibitemShut {NoStop}%
\bibitem [{\citenamefont {Moriya}(1960)}]{Moriya1960}%
  \BibitemOpen
  \bibfield  {author} {\bibinfo {author} {\bibfnamefont {T.}~\bibnamefont
  {Moriya}},\ }\href@noop {} {\bibfield  {journal} {\bibinfo  {journal}
  {Physical Review}\ }\textbf {\bibinfo {volume} {120}},\ \bibinfo {pages} {91}
  (\bibinfo {year} {1960})}\BibitemShut {NoStop}%
\bibitem [{\citenamefont {Yu}\ \emph {et~al.}(2011)\citenamefont {Yu},
  \citenamefont {Kanazawa}, \citenamefont {Onose}, \citenamefont {Kimoto},
  \citenamefont {Zhang}, \citenamefont {Ishiwata}, \citenamefont {Matsui},\
  and\ \citenamefont {Tokura}}]{Yu2011}%
  \BibitemOpen
  \bibfield  {author} {\bibinfo {author} {\bibfnamefont {X.~Z.}\ \bibnamefont
  {Yu}}, \bibinfo {author} {\bibfnamefont {N.}~\bibnamefont {Kanazawa}},
  \bibinfo {author} {\bibfnamefont {Y.}~\bibnamefont {Onose}}, \bibinfo
  {author} {\bibfnamefont {K.}~\bibnamefont {Kimoto}}, \bibinfo {author}
  {\bibfnamefont {W.~Z.}\ \bibnamefont {Zhang}}, \bibinfo {author}
  {\bibfnamefont {S.}~\bibnamefont {Ishiwata}}, \bibinfo {author}
  {\bibfnamefont {Y.}~\bibnamefont {Matsui}}, \ and\ \bibinfo {author}
  {\bibfnamefont {Y.}~\bibnamefont {Tokura}},\ }\href {\doibase
  https://doi.org/10.1038/nmat2916} {\bibfield  {journal} {\bibinfo  {journal}
  {Nature Materials}\ }\textbf {\bibinfo {volume} {12}},\ \bibinfo {pages}
  {106} (\bibinfo {year} {2011})}\BibitemShut {NoStop}%
\bibitem [{\citenamefont {Yu}\ \emph {et~al.}(2015)\citenamefont {Yu},
  \citenamefont {Kikkawa}, \citenamefont {Morikawa}, \citenamefont {Shibata},
  \citenamefont {Tokunaga}, \citenamefont {Taguchi},\ and\ \citenamefont
  {Tokura}}]{Yu2015}%
  \BibitemOpen
  \bibfield  {author} {\bibinfo {author} {\bibfnamefont {X.~Z.}\ \bibnamefont
  {Yu}}, \bibinfo {author} {\bibfnamefont {A.}~\bibnamefont {Kikkawa}},
  \bibinfo {author} {\bibfnamefont {D.}~\bibnamefont {Morikawa}}, \bibinfo
  {author} {\bibfnamefont {K.}~\bibnamefont {Shibata}}, \bibinfo {author}
  {\bibfnamefont {Y.}~\bibnamefont {Tokunaga}}, \bibinfo {author}
  {\bibfnamefont {Y.}~\bibnamefont {Taguchi}}, \ and\ \bibinfo {author}
  {\bibfnamefont {Y.}~\bibnamefont {Tokura}},\ }\href {\doibase
  https://doi.org/10.1103/PhysRevB.91.054411} {\bibfield  {journal} {\bibinfo
  {journal} {Physical Review B}\ }\textbf {\bibinfo {volume} {91}},\ \bibinfo
  {pages} {054411} (\bibinfo {year} {2015})}\BibitemShut {NoStop}%
\bibitem [{\citenamefont {Thiaville}\ \emph {et~al.}(2012)\citenamefont
  {Thiaville}, \citenamefont {Rohart}, \citenamefont {Jue}, \citenamefont
  {Cros},\ and\ \citenamefont {Fert}}]{Thiaville2012}%
  \BibitemOpen
  \bibfield  {author} {\bibinfo {author} {\bibfnamefont {A.}~\bibnamefont
  {Thiaville}}, \bibinfo {author} {\bibfnamefont {S.}~\bibnamefont {Rohart}},
  \bibinfo {author} {\bibfnamefont {E.}~\bibnamefont {Jue}}, \bibinfo {author}
  {\bibfnamefont {V.}~\bibnamefont {Cros}}, \ and\ \bibinfo {author}
  {\bibfnamefont {A.}~\bibnamefont {Fert}},\ }\href {<Go to
  ISI>://WOS:000312541700016
  http://iopscience.iop.org/0295-5075/100/5/57002/pdf/0295-5075_100_5_57002.pdf}
  {\bibfield  {journal} {\bibinfo  {journal} {EPL}\ }\textbf {\bibinfo {volume}
  {100}} (\bibinfo {year} {2012})}\BibitemShut {NoStop}%
\bibitem [{\citenamefont {Hrabec}\ \emph {et~al.}(2014)\citenamefont {Hrabec},
  \citenamefont {Porter}, \citenamefont {Wells}, \citenamefont {Benitez},
  \citenamefont {Burnell}, \citenamefont {McVitie}, \citenamefont {McGrouther},
  \citenamefont {Moore},\ and\ \citenamefont {Marrows}}]{Hrabec2014}%
  \BibitemOpen
  \bibfield  {author} {\bibinfo {author} {\bibfnamefont {A.}~\bibnamefont
  {Hrabec}}, \bibinfo {author} {\bibfnamefont {N.~A.}\ \bibnamefont {Porter}},
  \bibinfo {author} {\bibfnamefont {A.}~\bibnamefont {Wells}}, \bibinfo
  {author} {\bibfnamefont {M.~J.}\ \bibnamefont {Benitez}}, \bibinfo {author}
  {\bibfnamefont {G.}~\bibnamefont {Burnell}}, \bibinfo {author} {\bibfnamefont
  {S.}~\bibnamefont {McVitie}}, \bibinfo {author} {\bibfnamefont
  {D.}~\bibnamefont {McGrouther}}, \bibinfo {author} {\bibfnamefont {T.~A.}\
  \bibnamefont {Moore}}, \ and\ \bibinfo {author} {\bibfnamefont {C.~H.}\
  \bibnamefont {Marrows}},\ }\href {\doibase 10.1103/PhysRevB.90.020402}
  {\bibfield  {journal} {\bibinfo  {journal} {Phys. Rev. B}\ }\textbf {\bibinfo
  {volume} {90}},\ \bibinfo {pages} {020402} (\bibinfo {year}
  {2014})}\BibitemShut {NoStop}%
\bibitem [{\citenamefont {Pellegren}\ \emph {et~al.}(2017)\citenamefont
  {Pellegren}, \citenamefont {Lau},\ and\ \citenamefont
  {Sokalski}}]{Pellegren2017}%
  \BibitemOpen
  \bibfield  {author} {\bibinfo {author} {\bibfnamefont {J.~P.}\ \bibnamefont
  {Pellegren}}, \bibinfo {author} {\bibfnamefont {D.}~\bibnamefont {Lau}}, \
  and\ \bibinfo {author} {\bibfnamefont {V.}~\bibnamefont {Sokalski}},\ }\href
  {\doibase 10.1103/PhysRevLett.119.027203} {\bibfield  {journal} {\bibinfo
  {journal} {Phys. Rev. Lett.}\ }\textbf {\bibinfo {volume} {119}},\ \bibinfo
  {pages} {027203} (\bibinfo {year} {2017})}\BibitemShut {NoStop}%
\bibitem [{\citenamefont {Thiaville}(1995)}]{Thiaville1995}%
  \BibitemOpen
  \bibfield  {author} {\bibinfo {author} {\bibfnamefont {A.}~\bibnamefont
  {Thiaville}},\ }\href {\doibase https://doi.org/10.1016/0304-8853(94)00926-0}
  {\bibfield  {journal} {\bibinfo  {journal} {Journal of Magnetism and Magnetic
  Materials}\ }\textbf {\bibinfo {volume} {140-144}},\ \bibinfo {pages} {1877 }
  (\bibinfo {year} {1995})}\BibitemShut {NoStop}%
\bibitem [{\citenamefont {Chen}\ \emph {et~al.}(2013)\citenamefont {Chen},
  \citenamefont {Ma}, \citenamefont {N'Diaye}, \citenamefont {Kwon},
  \citenamefont {Won}, \citenamefont {Wu},\ and\ \citenamefont
  {Schmid}}]{Chen2013}%
  \BibitemOpen
  \bibfield  {author} {\bibinfo {author} {\bibfnamefont {G.}~\bibnamefont
  {Chen}}, \bibinfo {author} {\bibfnamefont {T.-P.}\ \bibnamefont {Ma}},
  \bibinfo {author} {\bibfnamefont {A.~T.}\ \bibnamefont {N'Diaye}}, \bibinfo
  {author} {\bibfnamefont {H.-Y.}\ \bibnamefont {Kwon}}, \bibinfo {author}
  {\bibfnamefont {C.-Y.}\ \bibnamefont {Won}}, \bibinfo {author} {\bibfnamefont
  {Y.-Z.}\ \bibnamefont {Wu}}, \ and\ \bibinfo {author} {\bibfnamefont {A.~K.}\
  \bibnamefont {Schmid}},\ }\href {\doibase https://doi.org/10.1038/ncomms3671}
  {\bibfield  {journal} {\bibinfo  {journal} {Nature Communications}\ }\textbf
  {\bibinfo {volume} {4}},\ \bibinfo {pages} {2671} (\bibinfo {year}
  {2013})}\BibitemShut {NoStop}%
\bibitem [{\citenamefont {Yang}\ \emph {et~al.}(2018)\citenamefont {Yang},
  \citenamefont {Boulle}, \citenamefont {Cros}, \citenamefont {Fert},\ and\
  \citenamefont {Chshiev}}]{Yang2018}%
  \BibitemOpen
  \bibfield  {author} {\bibinfo {author} {\bibfnamefont {H.-X.}\ \bibnamefont
  {Yang}}, \bibinfo {author} {\bibfnamefont {O.}~\bibnamefont {Boulle}},
  \bibinfo {author} {\bibfnamefont {V.}~\bibnamefont {Cros}}, \bibinfo {author}
  {\bibfnamefont {A.}~\bibnamefont {Fert}}, \ and\ \bibinfo {author}
  {\bibfnamefont {M.}~\bibnamefont {Chshiev}},\ }\href {\doibase
  10.1038/s41598-018-30063-y} {\bibfield  {journal} {\bibinfo  {journal}
  {Scientific Reports}\ ,\ \bibinfo {pages} {12356}} (\bibinfo {year}
  {2018})}\BibitemShut {NoStop}%
\bibitem [{\citenamefont {Khadka}\ \emph {et~al.}(2018)\citenamefont {Khadka},
  \citenamefont {Karayev},\ and\ \citenamefont {Huang}}]{Khadka2018}%
  \BibitemOpen
  \bibfield  {author} {\bibinfo {author} {\bibfnamefont {D.}~\bibnamefont
  {Khadka}}, \bibinfo {author} {\bibfnamefont {S.}~\bibnamefont {Karayev}}, \
  and\ \bibinfo {author} {\bibfnamefont {S.~X.}\ \bibnamefont {Huang}},\ }\href
  {\doibase 10.1063/1.5021090} {\bibfield  {journal} {\bibinfo  {journal}
  {Journal of Applied Physics}\ ,\ \bibinfo {pages} {123905}} (\bibinfo {year}
  {2018})}\BibitemShut {NoStop}%
\bibitem [{\citenamefont {Soumyanarayanan}\ \emph {et~al.}(2017)\citenamefont
  {Soumyanarayanan}, \citenamefont {Raju}, \citenamefont {Oyarce},
  \citenamefont {Tan}, \citenamefont {Im}, \citenamefont {Petrovi\'c},
  \citenamefont {Ho}, \citenamefont {Khoo}, \citenamefont {Tran}, \citenamefont
  {Gan}, \citenamefont {Ernult},\ and\ \citenamefont
  {Panagopoulos}}]{Soumy2017}%
  \BibitemOpen
  \bibfield  {author} {\bibinfo {author} {\bibfnamefont {A.}~\bibnamefont
  {Soumyanarayanan}}, \bibinfo {author} {\bibfnamefont {M.}~\bibnamefont
  {Raju}}, \bibinfo {author} {\bibfnamefont {A.~L.~G.}\ \bibnamefont {Oyarce}},
  \bibinfo {author} {\bibfnamefont {A.~K.~C.}\ \bibnamefont {Tan}}, \bibinfo
  {author} {\bibfnamefont {M.-Y.}\ \bibnamefont {Im}}, \bibinfo {author}
  {\bibfnamefont {A.~P.}\ \bibnamefont {Petrovi\'c}}, \bibinfo {author}
  {\bibfnamefont {P.}~\bibnamefont {Ho}}, \bibinfo {author} {\bibfnamefont
  {K.~H.}\ \bibnamefont {Khoo}}, \bibinfo {author} {\bibfnamefont
  {M.}~\bibnamefont {Tran}}, \bibinfo {author} {\bibfnamefont {C.~K.}\
  \bibnamefont {Gan}}, \bibinfo {author} {\bibfnamefont {F.}~\bibnamefont
  {Ernult}}, \ and\ \bibinfo {author} {\bibfnamefont {C.}~\bibnamefont
  {Panagopoulos}},\ }\href {\doibase https://doi.org/10.1038/nmat4934}
  {\bibfield  {journal} {\bibinfo  {journal} {Nature Materials}\ }\textbf
  {\bibinfo {volume} {16}},\ \bibinfo {pages} {898} (\bibinfo {year}
  {2017})}\BibitemShut {NoStop}%
\bibitem [{\citenamefont {Franken}\ \emph {et~al.}(2014)\citenamefont
  {Franken}, \citenamefont {Herps}, \citenamefont {Swagten},\ and\
  \citenamefont {Koopmans}}]{Franken2014}%
  \BibitemOpen
  \bibfield  {author} {\bibinfo {author} {\bibfnamefont {J.~H.}\ \bibnamefont
  {Franken}}, \bibinfo {author} {\bibfnamefont {M.}~\bibnamefont {Herps}},
  \bibinfo {author} {\bibfnamefont {H.~J.~M.}\ \bibnamefont {Swagten}}, \ and\
  \bibinfo {author} {\bibfnamefont {B.}~\bibnamefont {Koopmans}},\ }\href
  {\doibase https://doi.org/10.1038/srep05248} {\bibfield  {journal} {\bibinfo
  {journal} {Scientific Reports}\ }\textbf {\bibinfo {volume} {4}},\ \bibinfo
  {pages} {5248} (\bibinfo {year} {2014})}\BibitemShut {NoStop}%
\bibitem [{\citenamefont {Moreau-Luchaire}\ \emph {et~al.}(2016)\citenamefont
  {Moreau-Luchaire}, \citenamefont {Moutafis}, \citenamefont {Reyren},
  \citenamefont {Sampaio}, \citenamefont {Vaz}, \citenamefont {Horne},
  \citenamefont {Bouzehouane}, \citenamefont {Garcia}, \citenamefont
  {Deranlot}, \citenamefont {Warnicke}, \citenamefont {Wohlhuter},
  \citenamefont {George}, \citenamefont {Weigand}, \citenamefont {Raabe},
  \citenamefont {Cros},\ and\ \citenamefont {Fert}}]{MoreauLuchaire2016}%
  \BibitemOpen
  \bibfield  {author} {\bibinfo {author} {\bibfnamefont {C.}~\bibnamefont
  {Moreau-Luchaire}}, \bibinfo {author} {\bibfnamefont {C.}~\bibnamefont
  {Moutafis}}, \bibinfo {author} {\bibfnamefont {N.}~\bibnamefont {Reyren}},
  \bibinfo {author} {\bibfnamefont {J.}~\bibnamefont {Sampaio}}, \bibinfo
  {author} {\bibfnamefont {C.~A.~F.}\ \bibnamefont {Vaz}}, \bibinfo {author}
  {\bibfnamefont {N.~V.}\ \bibnamefont {Horne}}, \bibinfo {author}
  {\bibfnamefont {K.}~\bibnamefont {Bouzehouane}}, \bibinfo {author}
  {\bibfnamefont {K.}~\bibnamefont {Garcia}}, \bibinfo {author} {\bibfnamefont
  {C.}~\bibnamefont {Deranlot}}, \bibinfo {author} {\bibfnamefont
  {P.}~\bibnamefont {Warnicke}}, \bibinfo {author} {\bibfnamefont
  {P.}~\bibnamefont {Wohlhuter}}, \bibinfo {author} {\bibfnamefont {J.-M.}\
  \bibnamefont {George}}, \bibinfo {author} {\bibfnamefont {M.}~\bibnamefont
  {Weigand}}, \bibinfo {author} {\bibfnamefont {J.}~\bibnamefont {Raabe}},
  \bibinfo {author} {\bibfnamefont {V.}~\bibnamefont {Cros}}, \ and\ \bibinfo
  {author} {\bibfnamefont {A.}~\bibnamefont {Fert}},\ }\href {\doibase
  https://doi.org/10.1038/nnano.2015.313} {\bibfield  {journal} {\bibinfo
  {journal} {Nature Nanotechnology}\ }\textbf {\bibinfo {volume} {11}},\
  \bibinfo {pages} {444} (\bibinfo {year} {2016})}\BibitemShut {NoStop}%
\bibitem [{\citenamefont {{Lau}}\ \emph {et~al.}(2018)\citenamefont {{Lau}},
  \citenamefont {{Price Pellegren}}, \citenamefont {{Nembach}}, \citenamefont
  {{Shaw}},\ and\ \citenamefont {{Sokalski}}}]{Lau2018}%
  \BibitemOpen
  \bibfield  {author} {\bibinfo {author} {\bibfnamefont {D.}~\bibnamefont
  {{Lau}}}, \bibinfo {author} {\bibfnamefont {J.}~\bibnamefont {{Price
  Pellegren}}}, \bibinfo {author} {\bibfnamefont {H.}~\bibnamefont
  {{Nembach}}}, \bibinfo {author} {\bibfnamefont {J.}~\bibnamefont {{Shaw}}}, \
  and\ \bibinfo {author} {\bibfnamefont {V.}~\bibnamefont {{Sokalski}}},\
  }\href@noop {} {\bibfield  {journal} {\bibinfo  {journal} {ArXiv e-prints}\ }
  (\bibinfo {year} {2018})},\ \Eprint {http://arxiv.org/abs/1808.05520}
  {arXiv:1808.05520 [cond-mat.mtrl-sci]} \BibitemShut {NoStop}%
\bibitem [{\citenamefont {Je}\ \emph {et~al.}(2013)\citenamefont {Je},
  \citenamefont {Kim}, \citenamefont {Yoo}, \citenamefont {Min}, \citenamefont
  {Lee},\ and\ \citenamefont {Choe}}]{je2013}%
  \BibitemOpen
  \bibfield  {author} {\bibinfo {author} {\bibfnamefont {S.-G.}\ \bibnamefont
  {Je}}, \bibinfo {author} {\bibfnamefont {D.-H.}\ \bibnamefont {Kim}},
  \bibinfo {author} {\bibfnamefont {S.-C.}\ \bibnamefont {Yoo}}, \bibinfo
  {author} {\bibfnamefont {B.-C.}\ \bibnamefont {Min}}, \bibinfo {author}
  {\bibfnamefont {K.-J.}\ \bibnamefont {Lee}}, \ and\ \bibinfo {author}
  {\bibfnamefont {S.-B.}\ \bibnamefont {Choe}},\ }\href@noop {} {\bibfield
  {journal} {\bibinfo  {journal} {Physical Review B}\ }\textbf {\bibinfo
  {volume} {88}},\ \bibinfo {pages} {214401} (\bibinfo {year}
  {2013})}\BibitemShut {NoStop}%
\bibitem [{\citenamefont {Lavrijsen}\ \emph {et~al.}(2015)\citenamefont
  {Lavrijsen}, \citenamefont {Hartmann}, \citenamefont {van~den Brink},
  \citenamefont {Yin}, \citenamefont {Barcones}, \citenamefont {Duine},
  \citenamefont {Verheijen}, \citenamefont {Swagten},\ and\ \citenamefont
  {Koopmans}}]{lavrijsen2015}%
  \BibitemOpen
  \bibfield  {author} {\bibinfo {author} {\bibfnamefont {R.}~\bibnamefont
  {Lavrijsen}}, \bibinfo {author} {\bibfnamefont {D.~M.~F.}\ \bibnamefont
  {Hartmann}}, \bibinfo {author} {\bibfnamefont {A.}~\bibnamefont {van~den
  Brink}}, \bibinfo {author} {\bibfnamefont {Y.}~\bibnamefont {Yin}}, \bibinfo
  {author} {\bibfnamefont {B.}~\bibnamefont {Barcones}}, \bibinfo {author}
  {\bibfnamefont {R.~A.}\ \bibnamefont {Duine}}, \bibinfo {author}
  {\bibfnamefont {M.~A.}\ \bibnamefont {Verheijen}}, \bibinfo {author}
  {\bibfnamefont {H.~J.~M.}\ \bibnamefont {Swagten}}, \ and\ \bibinfo {author}
  {\bibfnamefont {B.}~\bibnamefont {Koopmans}},\ }\href {\doibase
  10.1103/PhysRevB.91.104414} {\bibfield  {journal} {\bibinfo  {journal} {Phys.
  Rev. B}\ }\textbf {\bibinfo {volume} {91}},\ \bibinfo {pages} {104414}
  (\bibinfo {year} {2015})}\BibitemShut {NoStop}%
\bibitem [{Sup()}]{SupMat}%
  \BibitemOpen
  \href@noop {} {}\bibinfo {howpublished} {See Supplemental Material at [URL
  will be inserted by publisher] for measurement of DMI and additional Lorentz
  TEM images, which includes \cite{Hrabec2014, Lau2016, Pellegren2017,
  Lau2018}}\BibitemShut {NoStop}%
\bibitem [{\citenamefont {KOOY}(1960)}]{Kooy1960}%
  \BibitemOpen
  \bibfield  {author} {\bibinfo {author} {\bibfnamefont {C.}~\bibnamefont
  {KOOY}},\ }\href {https://ci.nii.ac.jp/naid/10006485428/en/} {\bibfield
  {journal} {\bibinfo  {journal} {Philips Res. Repts}\ }\textbf {\bibinfo
  {volume} {15}} (\bibinfo {year} {1960})}\BibitemShut {NoStop}%
\bibitem [{\citenamefont {Cape}\ and\ \citenamefont {Lehman}(1971)}]{Cape1971}%
  \BibitemOpen
  \bibfield  {author} {\bibinfo {author} {\bibfnamefont {J.~A.}\ \bibnamefont
  {Cape}}\ and\ \bibinfo {author} {\bibfnamefont {G.~W.}\ \bibnamefont
  {Lehman}},\ }\href {\doibase 10.1063/1.1660007} {\bibfield  {journal}
  {\bibinfo  {journal} {Journal of Applied Physics}\ }\textbf {\bibinfo
  {volume} {42}},\ \bibinfo {pages} {5732} (\bibinfo {year} {1971})},\ \Eprint
  {http://arxiv.org/abs/https://doi.org/10.1063/1.1660007}
  {https://doi.org/10.1063/1.1660007} \BibitemShut {NoStop}%
\bibitem [{\citenamefont {Phatak}\ \emph {et~al.}(2016)\citenamefont {Phatak},
  \citenamefont {Heinonen}, \citenamefont {De~Graef},\ and\ \citenamefont
  {Petford-Long}}]{phatak2016}%
  \BibitemOpen
  \bibfield  {author} {\bibinfo {author} {\bibfnamefont {C.}~\bibnamefont
  {Phatak}}, \bibinfo {author} {\bibfnamefont {O.}~\bibnamefont {Heinonen}},
  \bibinfo {author} {\bibfnamefont {M.}~\bibnamefont {De~Graef}}, \ and\
  \bibinfo {author} {\bibfnamefont {A.}~\bibnamefont {Petford-Long}},\
  }\href@noop {} {\bibfield  {journal} {\bibinfo  {journal} {Nano letters}\
  }\textbf {\bibinfo {volume} {16}},\ \bibinfo {pages} {4141} (\bibinfo {year}
  {2016})}\BibitemShut {NoStop}%
\bibitem [{\citenamefont {Benitez}\ \emph {et~al.}(2015)\citenamefont
  {Benitez}, \citenamefont {Hrabec}, \citenamefont {Mihai}, \citenamefont
  {Moore}, \citenamefont {Burnell}, \citenamefont {McGrouther}, \citenamefont
  {Marrows},\ and\ \citenamefont {McVitie}}]{Benitez2015}%
  \BibitemOpen
  \bibfield  {author} {\bibinfo {author} {\bibfnamefont {M.~J.}\ \bibnamefont
  {Benitez}}, \bibinfo {author} {\bibfnamefont {A.}~\bibnamefont {Hrabec}},
  \bibinfo {author} {\bibfnamefont {A.~P.}\ \bibnamefont {Mihai}}, \bibinfo
  {author} {\bibfnamefont {T.~A.}\ \bibnamefont {Moore}}, \bibinfo {author}
  {\bibfnamefont {G.}~\bibnamefont {Burnell}}, \bibinfo {author} {\bibfnamefont
  {D.}~\bibnamefont {McGrouther}}, \bibinfo {author} {\bibfnamefont {C.~H.}\
  \bibnamefont {Marrows}}, \ and\ \bibinfo {author} {\bibfnamefont
  {S.}~\bibnamefont {McVitie}},\ }\href {http://dx.doi.org/10.1038/ncomms9957}
  {\bibfield  {journal} {\bibinfo  {journal} {Nature Communications}\ }\textbf
  {\bibinfo {volume} {6}},\ \bibinfo {pages} {8957 EP } (\bibinfo {year}
  {2015})}\BibitemShut {NoStop}%
\bibitem [{\citenamefont {McVitie}\ \emph {et~al.}(2018)\citenamefont
  {McVitie}, \citenamefont {Hughes}, \citenamefont {Fallon}, \citenamefont
  {McFadzean}, \citenamefont {McGrouther}, \citenamefont {Krajnak},
  \citenamefont {Legrand}, \citenamefont {Maccariello}, \citenamefont {Collin},
  \citenamefont {Garcia}, \citenamefont {Reyren}, \citenamefont {Cros},
  \citenamefont {Fert}, \citenamefont {Zeissler},\ and\ \citenamefont
  {Marrows}}]{McVitie2018}%
  \BibitemOpen
  \bibfield  {author} {\bibinfo {author} {\bibfnamefont {S.}~\bibnamefont
  {McVitie}}, \bibinfo {author} {\bibfnamefont {S.}~\bibnamefont {Hughes}},
  \bibinfo {author} {\bibfnamefont {K.}~\bibnamefont {Fallon}}, \bibinfo
  {author} {\bibfnamefont {S.}~\bibnamefont {McFadzean}}, \bibinfo {author}
  {\bibfnamefont {D.}~\bibnamefont {McGrouther}}, \bibinfo {author}
  {\bibfnamefont {M.}~\bibnamefont {Krajnak}}, \bibinfo {author} {\bibfnamefont
  {W.}~\bibnamefont {Legrand}}, \bibinfo {author} {\bibfnamefont
  {D.}~\bibnamefont {Maccariello}}, \bibinfo {author} {\bibfnamefont
  {S.}~\bibnamefont {Collin}}, \bibinfo {author} {\bibfnamefont
  {K.}~\bibnamefont {Garcia}}, \bibinfo {author} {\bibfnamefont
  {N.}~\bibnamefont {Reyren}}, \bibinfo {author} {\bibfnamefont
  {V.}~\bibnamefont {Cros}}, \bibinfo {author} {\bibfnamefont {A.}~\bibnamefont
  {Fert}}, \bibinfo {author} {\bibfnamefont {K.}~\bibnamefont {Zeissler}}, \
  and\ \bibinfo {author} {\bibfnamefont {C.~H.}\ \bibnamefont {Marrows}},\
  }\href {\doibase 10.1038/s41598-018-23799-0} {\bibfield  {journal} {\bibinfo
  {journal} {Scientific Reports}\ }\textbf {\bibinfo {volume} {8}},\ \bibinfo
  {pages} {5703} (\bibinfo {year} {2018})}\BibitemShut {NoStop}%
\bibitem [{\citenamefont {Malozemoff}\ and\ \citenamefont
  {Slonczewski}(1972)}]{malozemoff1972}%
  \BibitemOpen
  \bibfield  {author} {\bibinfo {author} {\bibfnamefont {A.}~\bibnamefont
  {Malozemoff}}\ and\ \bibinfo {author} {\bibfnamefont {J.}~\bibnamefont
  {Slonczewski}},\ }\href@noop {} {\bibfield  {journal} {\bibinfo  {journal}
  {Physical Review Letters}\ }\textbf {\bibinfo {volume} {29}},\ \bibinfo
  {pages} {952} (\bibinfo {year} {1972})}\BibitemShut {NoStop}%
\bibitem [{\citenamefont {Zvezdin}\ and\ \citenamefont
  {Popkov}(1986)}]{zvezdin1986}%
  \BibitemOpen
  \bibfield  {author} {\bibinfo {author} {\bibfnamefont {A.}~\bibnamefont
  {Zvezdin}}\ and\ \bibinfo {author} {\bibfnamefont {A.}~\bibnamefont
  {Popkov}},\ }\href@noop {} {\bibfield  {journal} {\bibinfo  {journal} {Zh.
  Eksp. Teor. Fiz}\ }\textbf {\bibinfo {volume} {91}},\ \bibinfo {pages} {1789}
  (\bibinfo {year} {1986})}\BibitemShut {NoStop}%
\bibitem [{\citenamefont {Lau}\ \emph {et~al.}(2016)\citenamefont {Lau},
  \citenamefont {Sundar}, \citenamefont {Zhu},\ and\ \citenamefont
  {Sokalski}}]{Lau2016}%
  \BibitemOpen
  \bibfield  {author} {\bibinfo {author} {\bibfnamefont {D.}~\bibnamefont
  {Lau}}, \bibinfo {author} {\bibfnamefont {V.}~\bibnamefont {Sundar}},
  \bibinfo {author} {\bibfnamefont {J.-G.}\ \bibnamefont {Zhu}}, \ and\
  \bibinfo {author} {\bibfnamefont {V.}~\bibnamefont {Sokalski}},\ }\href
  {\doibase 10.1103/PhysRevB.94.060401} {\bibfield  {journal} {\bibinfo
  {journal} {Phys. Rev. B}\ }\textbf {\bibinfo {volume} {94}},\ \bibinfo
  {pages} {060401} (\bibinfo {year} {2016})}\BibitemShut {NoStop}%
\end{thebibliography}%

\end{document}